\documentclass[12pt,preprint]{aastex}

\shorttitle{The identification of IGR J17204$-$3554}
\shortauthors{Bassani et al.}

\begin{document}


\title{Is the INTEGRAL/IBIS source IGR J17204$-$3554 a $\gamma$-ray emitting galaxy
hidden behind the molecular cloud NGC 6334 ?\altaffilmark{1}}

\author{L.~Bassani\altaffilmark{2}, A.~De~Rosa\altaffilmark{3}, A.~Bazzano\altaffilmark{3},
A.~J.~Bird\altaffilmark{4}, A.~J.~Dean\altaffilmark{4}, N.~Gehrels\altaffilmark{5},
J.~A.~Kennea\altaffilmark{6}, A.~Malizia\altaffilmark{2}, M.~Molina\altaffilmark{4}, J.~B.~Stephen\altaffilmark{2},
P.~Ubertini\altaffilmark{3}, R.~Walter\altaffilmark{7}}
\altaffiltext{1}{Based on observations obtained with the ESA science mission {\it INTEGRAL}}.
\altaffiltext{2}{IASF-Bologna/INAF, Via Gobetti 101, I-40129 Bologna, Italy}
\altaffiltext{3}{IASF-Roma/INAF, Via Fosso del Cavaliere 100, I-00133 Rome, Italy} 
\altaffiltext{4}{School of Physics and Astronomy, University of
Southampton, Highfield, Southampton, SO 17 1BJ, UK.}
\altaffiltext{5}{NASA/Goddard Space Flight Center, Greenbelt, MD 20771}
\altaffiltext{6}{Department of Astronomy and Astrophysics, 525 Davey Lab, Pennsylvania Sate 
University, University Park, PA 16802}
\altaffiltext{7}{{\it INTEGRAL} Science Data Centre,Chemin d'Ecogia 16,1291, Versoix, Switzerland}

\begin{abstract}
We report on the identification of a soft gamma-ray source, IGR J17204$-$3554, detected with the IBIS imager on board the INTEGRAL satellite. The source has a 20-100 keV flux of $\sim$3$\times$10$^{-11}$ erg cm$^{-2}$ s$^{-1}$ and is spatially coincident with NGC 6334, a molecular cloud located in the Sagittarius arm of the Milky Way.  Diffuse X-ray emission has been reported from this region by ASCA and interpreted as coming from five far-infrared cores located in the cloud. However, the combined ASCA spectrum with a 9 keV temperature was difficult to explain in terms of emission from young pre-main sequence stars known to be embedded in the star forming regions. Detection of gamma-rays makes this interpretation even more unrealistic and suggests the presence of a high energy source in or behind the cloud.  Follow up observations with Swift and archival Chandra data allow us to disentangle the NGC6334 enigma by locating an extragalactic object with the proper spectral characteristics to explain the $\gamma$-ray emission.  The combined Chandra/IBIS spectrum is well fitted by an absorbed power law with $\Gamma$=1.2$\pm$0.1, N$_{H}$=1.4$\pm$0.1$\times$10$^{23}$ cm$^{-2}$ and an unabsorbed 2-10 keV flux of 0.5$\times$10$^{-11}$ erg cm$^{-2}$ s$^{-1}$.  This column density is in excess of the galactic value implying that we are detecting a background galaxy concealed by the molecular cloud and further hidden by material located either in the galaxy itself or between IGR J17204$-$3554 and the cloud.  
\end{abstract}


\section{Introduction}
Giant Molecular Clouds (GMC) are the coolest (10-20K) and densest
portions (about 10$^{12}$ particles per cubic meter) of the interstellar
medium: stretching typically over 100 light-years and containing
several hundred thousand solar masses of material, they are the
largest known objects in the Universe made of molecular material.
Observations of these sky regions are made difficult by the large
amount of gas and dust which prevents direct optical view. The only
source of information on objects inside or behind GMCs is provided at
longer wavelengths such as radio and infrared where the emission is
free from absorption.  Although high energy measurements can also be
extremely efficient in probing deeply into these regions, only a
handful of GMC have so far been observed in X-rays (Garmire et al. 2000,
Hofner et al. 2002, Kohno, Koyama \& Hamaguchi 2002) and virtually none
in gamma-rays.  To probe GMC regions in gamma-rays, the 
imager on board INTEGRAL (IBIS) is a powerful instrument: it allows source
detection above 20 keV with a mCrab ($\sim $10$^{-11}$ erg cm$^{-2}$
s$^{-1}$) sensitivity in well exposed regions, an angular resolution
of 12\arcmin (thus covering the full extension of a GMC) and a point
source location accuracy of 1\arcmin-2\arcmin for moderately bright sources
(Ubertini et al. 2003). Furthermore, INTEGRAL has regularly observed
the entire galactic plane during the first two and half years in orbit
providing, at these energies, a galactic survey with unprecedented
sensitivity (Bird et al. 2004). A second catalogue, resulting from greater
sky coverage and deeper exposures, is now completed (Bird et
al.  2005): within this catalogue two sources are spatially coincident
with molecular clouds, IGR J17475$-$2822 and IGR J17204$-$3554.  The first
object is fully discussed by Revnivtsev et al. (2004) who interpreted
the X/soft-gamma ray emission as Compton scattered and reprocessed
radiation emitted in the past by our Galaxy Center (Sgr A$^{\star}$)
and mirrored by the SGR B2 molecular cloud complex.  In this paper, we
report, instead, on the identification of the second object which
is associated with NGC6334. Our deep analysis of the region indicates
that this new IBIS source is probably a background AGN seen through
the cloud.

\section{NGC6334}
NGC6334 is a prominent HII region/molecular cloud complex located in
the Sagittarius arm of the Milky Way at a photometric distance of 1.7
kpc (Neckel 1978).  The complex contains
several recent and current star-forming sites which are embedded in an
elongated GMC extending over about 45\arcmin
(Dickel, Dickel \& Wilson 1977; Kraemer \& Jackson 1999).  The 
morphology of the region is complex, and, unfortunately, this is reflected in
the nomenclature of the different sources and components. In the
scheme that has evolved in time, letter designations correspond to
centimeter radio sources, while roman numeral designations 
are used for millimeter and infrared sources.  Radio continuum
observations have identified five major components denoted as NGC 6334
A to F (Rodr\'{i}guez, Cant\'{o}, \& Moran 1982) while far-infrared (FIR)
observations have detected six major sources that are denoted NGC 6334
I to VI, increasing in the opposite direction relative to the radio
data (McBreen et al. 1979).  In radio, all sources are HII regions,
with the exception of NGC6334B, which is probably an extragalactic
object seen through the cloud (Moran et al. 1990).  Each of the
far-infrared cores is instead due to the combined emission of young
massive stars embedded in the cloud star-forming regions (Rodr\'{i}guez,
Cant\'{o}, \& Moran 1982).  Although most of the reported cores are
detected both at radio and far-infrared wavelengths, not all radio
sources have corresponding FIR emission and vice versa (Kraemer \&
Jackson 1999).

X-ray emission has been detected by ASCA from NGC6334
(Sekimoto et al.  2000). Due to the limited angular resolution ($\ge$ 4\arcmin)
of the instrument it was not possible to separate the blurred extended emission
into single sources although five FIR cores were indicated as being responsible
for the emission above 2 keV (NGC6334 I to V) while at softer energies
the radiation was found to be mostly absorbed except for core III, located
at the center of the cloud.  Northwest of NGC6334 a bright object was
also detected: this corresponds to an emission line star of type BO.5 
(CD$-$3511482).
The combined ASCA spectrum was reported to be thermal with a temperature of $\sim$9
keV, a metal abundance about half the solar value, a column density of 9$\times$10$^{21}$ 
cm$^{-2}$ (close to the galactic value in the source
direction) and an unabsorbed flux of $\sim$2$\times$ 10$^{-11}$
erg cm$^{-2}$ s$^{-1}$ in the 0.5-10 keV band. The high temperature
observed is not easily reconciled with emission from young stellar
objects which are known to populate the cloud, so that alternative
explanations were proposed by the authors, none of which were convincing
enough to explain the high temperature seen by ASCA.

\section{Unveiling the Nature of IGR J17204$-$3554}
\subsection{Step 1: INTEGRAL/ASCA data}
Data reported here belong to the Core Program (i.e. were collected as
part of the INTEGRAL Galactic Plane Survey and Galactic Center Deep
Exposure) as well as to public Open Time observations and span from
revolution 46 (February 2003)  to revolution 202 (September 2004); 
the total exposure of the region containing IGR J17204$-$3554 is
1.36 Msec.  In the present paper, we refer to data collected by the
imager (IBIS) on board INTEGRAL (Ubertini et al. 2003) and in
particular to detection by the first layer (ISGRI) of the instrument
(Lebrun et al. 2003).  A detailed description of the source extraction
criteria can be found in Bird et al. (2005).  Figure 1 shows the
20-100 keV band image of the region surrounding IGR J17204$-$3554: the
source is detected with a significance of $\sim$13$\sigma$ 
at a position corresponding to R.A.(2000)=17$^{\mathrm h}$20$^{\mathrm m}$24\fs96
and Dec(2000)=-35\degr54\arcmin00\farcs00 with a positional uncertainty
of $\leq$1\farcm5 (90\% confidence, Gros et al. 2003).
Superposition of the INTEGRAL/IBIS positional uncertainty on the ASCA/GIS
image (figure 2) indicates that the IBIS emission is located between
infrared Cores III and IV. Given the morphology of the
region, we cannot exclude at this stage that we are detecting diffuse
and/or multiple source emission.

In view of  its angular resolution, IBIS/ISGRI is not able to separate the various contributions:
in this case the emission from the whole region, including the stellar
object at the northwest side of the cloud, is detected by INTEGRAL.
The flux of IGR J17204$-$3554 detected in each individual pointing was
also used to generate the source light curve: no flares are visible
nor does the source show periodicities and/or pulsations.
Fluxes for spectral analysis were extracted from narrow band mosaics of
all revolutions added together.  A simple power law provides a good
fit to the IBIS data ($\chi^{2}$=6.4/4), a flat photon index
($\Gamma$=1.43$^{+0.26}_{-0.25}$) and a 20-100 keV flux of 3$\times$10$^{-11}$ 
erg cm$^{-2}$ s$^{-1}$.  Quoted errors here and in the
following correspond to 90${\%}$ confidence level for one interesting parameter.

Next we analysed the combined ASCA/GIS and INTEGRAL/IBIS data over the
2-100 keV band; the ASCA data refer to a region of 4\arcmin centered on
the cloud.  The combined data are well fitted ($\chi^{2}$=162.5/143) by an absorbed power law
having a photon index of $\Gamma$=1.43$^{+0.05}_{-0.04}$ and a column
density of N$_{H}$=7.5$\pm$0.9$\times$10$^{21}$ cm$^{-2}$. To
account for a cross calibration mismatch between the two instruments
and/or source variability between the two observing periods, we have
introduced a free constant in the fit; when left free to vary it
provides a value of 1.3$\pm$0.3.  A thermal model could
also be a good fit but results in an unacceptably high temperature of
130 keV; if the temperature is constrained to the value originally proposed by ASCA, 
the fit is unacceptable implying that a thermal model is
unable to explain the soft gamma-ray emission detected by INTEGRAL.

\subsection{Step 2: Swift/Chandra data}
In order to understand what powers the gamma-rays seen from
NGC6334 by INTEGRAL, a ToO observation with Swift (Gehrels et al.
2004) was immediately requested and granted.  The observation was
performed on July 12, 2005 and the source observed for 1.5 ks  with the XRT in Photon Counting 
mode (Hill et al. 2004). Data
reduction was performed using the XRTDAS v2.0 standard data pipeline package.  Two sources
were readily detected in the region: source 1 at R.A.(2000)=17$^{\mathrm h}$20$^{\mathrm m}$31\fs88 and 
Dec(2000)=-35\degr51\arcmin04\farcs63 and source 2 at
R.A.(2000)=17$^{\mathrm h}$20$^{\mathrm m}$26\fs00 and Dec(2000)=-35\degr43\arcmin58\farcs80 with
a positional uncertainty of 7 and 10 arcsec respectively.  Both
sources are quite weak in X-rays with 0.5-10 keV fluxes of 2.7 and 1.3
$\times$10$^{-12}$ erg cm$^{-2}$ s$^{-1}$ respectively.  The first
source (also named 2E1717.1-3548) coincides with [SHM89] FIR-III-13, a zero age main sequence star of 
type O7 which is thought to ionize the HII region NGC 6334C/Core III (Straw, Hayland $\&$ McGregor 1989);
the second is coincident with the emission line star CD-3511482 also seen by ASCA.
Only source 1 is close to the ISGRI
error box, although outside its border but it is too weak to be
associated with the ISGRI source unless it is strongly
absorbed.  However, this is the source associated with core III which
is the only one seen by ASCA at soft energies suggesting therefore
weak absorbtion. Despite the interesting result of pinpointing X-ray
emission from the star ionizing a compact HII region, the Swift observation was too short to be able 
to solve the ISGRI source enigma and to answer the question of what is
producing gamma-rays in a molecular cloud.  

A search of the HEASARC
archive provided 2 Chandra observations on 31-08-2002 (OP1) and 02-09-2002 (OP2). 
These measurements,  made with the ACIS instrument, lasted  40 ks
each and were pointed at R.A.(2000)=17$^{\mathrm h}$20$^{\mathrm m}$01$^{\mathrm s}$, Dec(2000)=-35\degr56\arcmin07\arcsec (OP1) and 
R.A.(2000)=17$^{\mathrm h}$20$^{\mathrm m}$54$^{\mathrm s}$, Dec(2000)=-35\degr47\arcmin04\arcsec (OP2) respectively. They each cover a portion of
the cloud and OP1 in particular also completely covered the
ISGRI error box.  The Chandra data were reduced following standard
procedures and using CIAO v3.2.  A quick-look analysis of these data
indicates that the emission in the region of interest is resolved
into many point sources, including source 1 (seen in  OP1/OP2 and 
shown in figure 3) and source 2
(seen in OP1 only). The Chandra position and location accuracy
($\le$1\arcsec) confirm the Swift identification for source 1 and further provide information on sources 
inside  the ISGRI error box: many X-ray emitting objects are detected, but they are  too X-ray weak for 
detection by ISGRI except for source 3
located at R.A.(2000)=17$^{\mathrm h}$20$^{\mathrm m}$21\fs81 and Dec(2000)=-35\degr52\arcmin48\farcs25 
(uncertainty $\le$1").  Figure 3, which shows
Chandra images in two different wavebands (0.5-2keV and above 2 keV),
clearly indicates that this source is very hard (heavily absorbed
and/or spectrally flat) as it is not seen below 2 keV but it is quite
bright above this energy compared to source 1 which is visible in both
wavebands.  In fact spectral analysis of Chandra data indicates that
source 1 is very soft and only slightly absorbed ($\Gamma$=2.36$\pm$0.12,
N$_{H}$=7.4$\pm$0.8$\times$10$^{21}$ cm$^{-2}$, $\chi^{2}$=147/123
d.o.f.) while source 3 is flat and heavily absorbed
($\Gamma$=1.6$\pm$0.4, N$_{H}$=1.6$\pm$0.3$\times$10$^{23}$
cm$^{-2}$, $\chi^{2}$=97/101).  This was the clue to the solution of the knotty
problem of NGC6334 by identifying source 3 as that responsible for
the soft gamma-ray emission.  In fact a combined analysis of ISGRI and
Chandra data of this source provides a good fit ($\chi^{2}$=109/107,
figure 4 ) with the following spectral parameters: $\Gamma$=1.2$\pm$0.1, 
N$_{H}$=1.4$\pm$0.1$\times$10$^{23}$ cm$^{-2}$ and a cross
calibration constant fixed to 1; if left free to vary
the constant is 1.8$^{+0.8}_{-0.6}$ and 
the spectrum is softer (1.5$\pm$0.3) while the absorption remains the same. 
The  flux corrected for absorption is 0.5$\times$10$^{-11}$ erg cm$^{-2}$ s$^{-1}$ in the 2-10 band. 

\subsection{Step 3: The nature of source 3}
Having found the source responsible for the X/gamma-ray emission in NGC6334, 
we next need to understand its nature.  The
Chandra position coincides with NGC6334B (also G351.28+0.68), the only
radio source possibly not associated with the molecular cloud but likely
to be a background active galaxy. This conclusion is based on a number
of observations: its brightness temperature at 6cm is
far in excess of the value expected from HII regions; it is not
closely associated with other sources of star formation (in fact it is not a FIR core); 
it has the largest scattering disk of any known source (implying plasma scattering
from the nearby NGC6334A HII region) and it is time variable at radio
frequencies (Moran et al. 1990). Also, measurements of HI indicate the presence of
additional absorption toward  NGC6334B with respect to the cloud 
and therefore a larger distance ($\ge$ 6 Kpc, Moran et al. 1990). 

Within the Chandra
error box there is no optical/infrared source listed in the USNO B1 or 2MASS
catalogues (Monet et al. 2003, Cutri et al. 2003), although there is slight evidence of very faint emission (magnitude $\sim$15) in the 2MASS K-band image. Correcting for measured X-ray absorption, this corresponds to an intrinsic brightness of 7$^{th}$ magnitude, similar to active galaxies in the local universe (D $\le$ 15 Mpc). Clearly follow up optical spectroscopy will 
better classify the source type and redshift, however, the extra absorption found 
suggests that it might be a type 2 active galaxy. Alternatively, the column  density in excess to the cloud value
might be related to material located between the cloud and the galaxy. In either case, NGC6334B
is a source concealed behind the molecular cloud and further hidden from
view by  extra gas and dust : a real conspiracy to prevent its detection.

\section{Conclusions}

Following  a difficult and  challenging path, 
we were able to identify  the newly discovered source, 
IGR J17204-3554, as a background galaxy.
While revealing the true nature of this IBIS/ISGRI object, we have also solved some of the
mysteries related to the X-ray emission from NGC6334. First we
can firmly state that the radiation measured by ASCA is resolved into
many weak point sources likely to be young massive stars embedded in
their forming regions as already observed in a number of molecular
clouds  (Garmire et al. 2000, Hofner et al. 2002, Kohno, Koyama and Hamaguchi 2002). 
Analysis of their X-ray characteristics is beyond the scope of the present paper
but already we can assess that they cannot provide the 9 keV
temperature reported by ASCA.  Instead this is due to the convolution
of many soft X-ray spectra contaminated at high energies by the emission  from IGR
J17204-3554. We can further conclude that the X-ray emission from NGC6334 is dominated
by 3 bright sources : an emission line star (CD-3511482), the ionizing
star of the ultracompact HII region NGC6334C ([SHM89] FIR-III-13) and an active galaxy
located behind the molecular cloud.  The radiation from this active
galaxy, besides being hidden by the cloud, is also absorbed by  material  located in the galaxy itself 
or in between IGR J17204-3554 and the cloud. Clearly, everything
was conspiring to prevent identification of this soft gamma-ray source and only 
the combination of 3 powerful instruments (INTEGRAL,
Swift and Chandra)
has allowed  us to finally solve the NGC6334 enigma.\\

\acknowledgments 
We acknowledge the following funding agencies: in Italy, the Italian Space Agency  
support through contract I/R/046/04; in the UK, 
the Particle Physics and Astronomy Research Council help via grant GR/2002/00446. 
Work at Penn State University
was supported by NASA under contract NAS5-00136. 
We thank P. Persi for useful discussions on the molecular cloud complex.
This research has made use of the SIMBAD database, operated at CDS, Strasbourg, France
and of the HEASARC archive provided by NASA's Goddard Space Flight Center.

\clearpage

\begin{figure}
\plotone{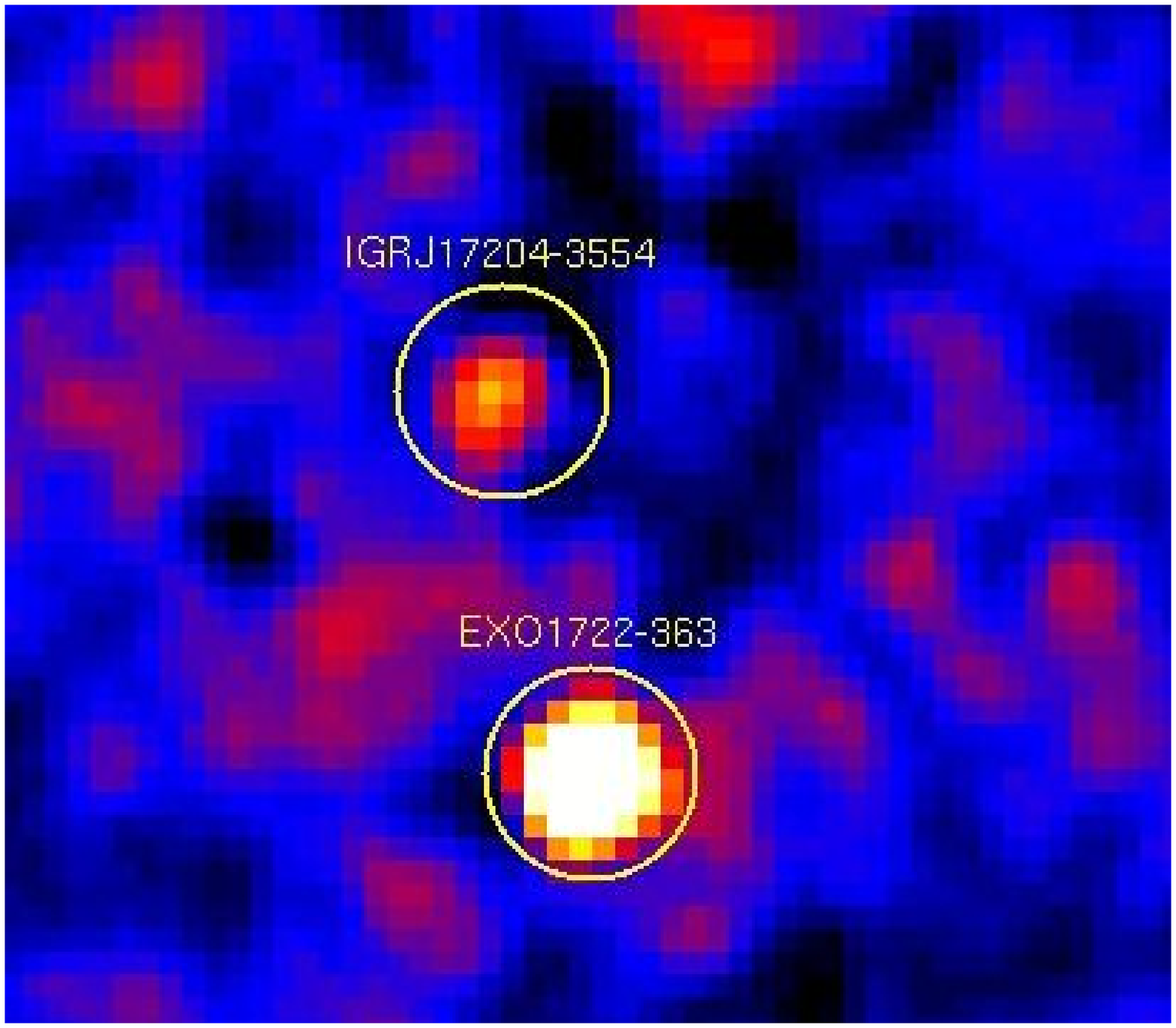} \caption {IBIS/ISGRI 20-100 keV image with 2 sources detected:
the HMXB EXO 1722$-$363 and the unidentified source IGR J17204$-$3554. \label{fig1}}
\end{figure}

\clearpage

\begin{figure}
\plotone{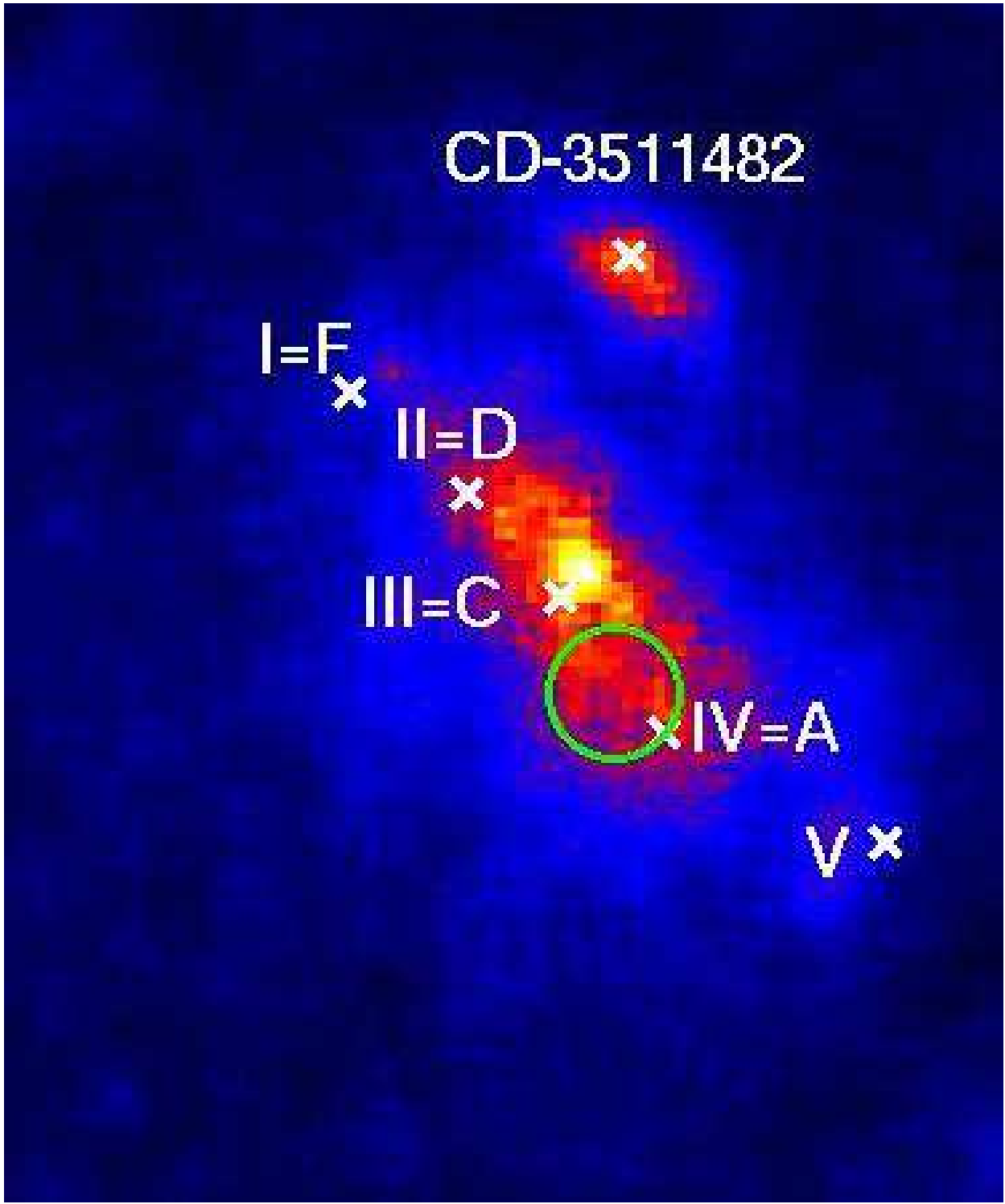} \caption {ASCA/GIS (2-10 keV) image: crosses mark the
positions of FIR cores I to V and corresponding radio sources. The circle corresponds to the IBIS/ISGRI error box. 
 \label{fig2}}
\end{figure}

\clearpage

\begin{figure}
\plotone{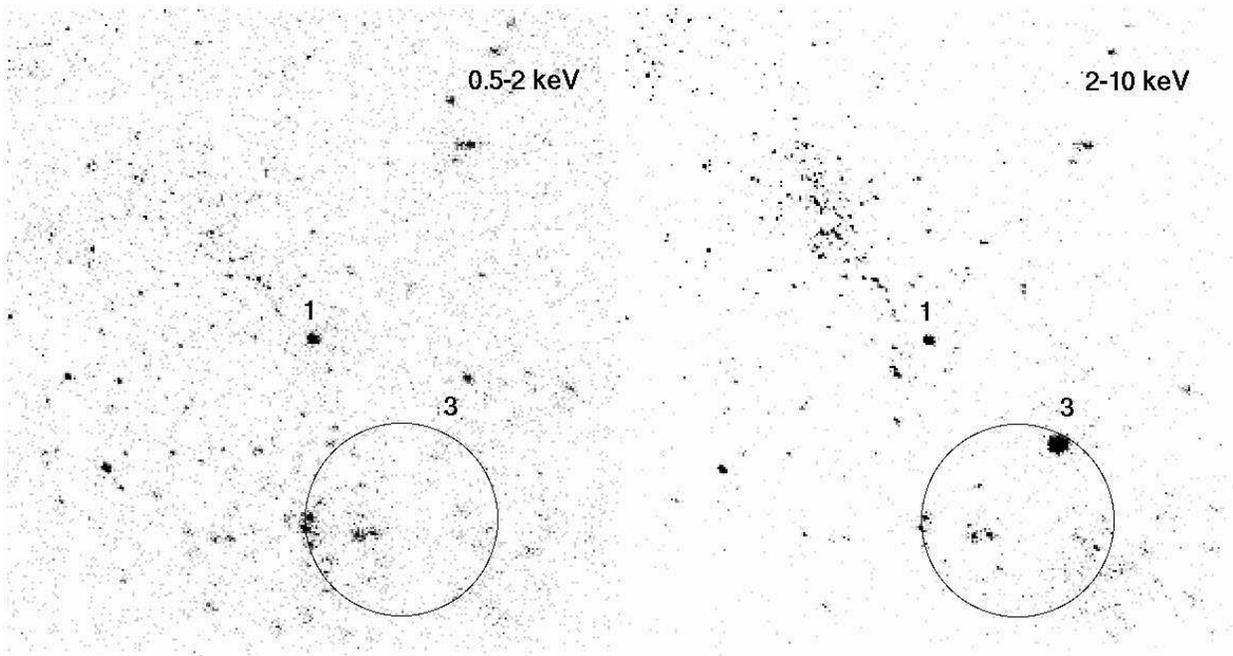} \caption {Chandra images in two energy bands.
The circle indicates the ISGRI error box.  Source 1, seen also by Swift, 
is soft and unabsorbed while source 3 is hard and absorbed. \label{fig3}}
\end{figure}

\clearpage

\begin{figure}
\plotone{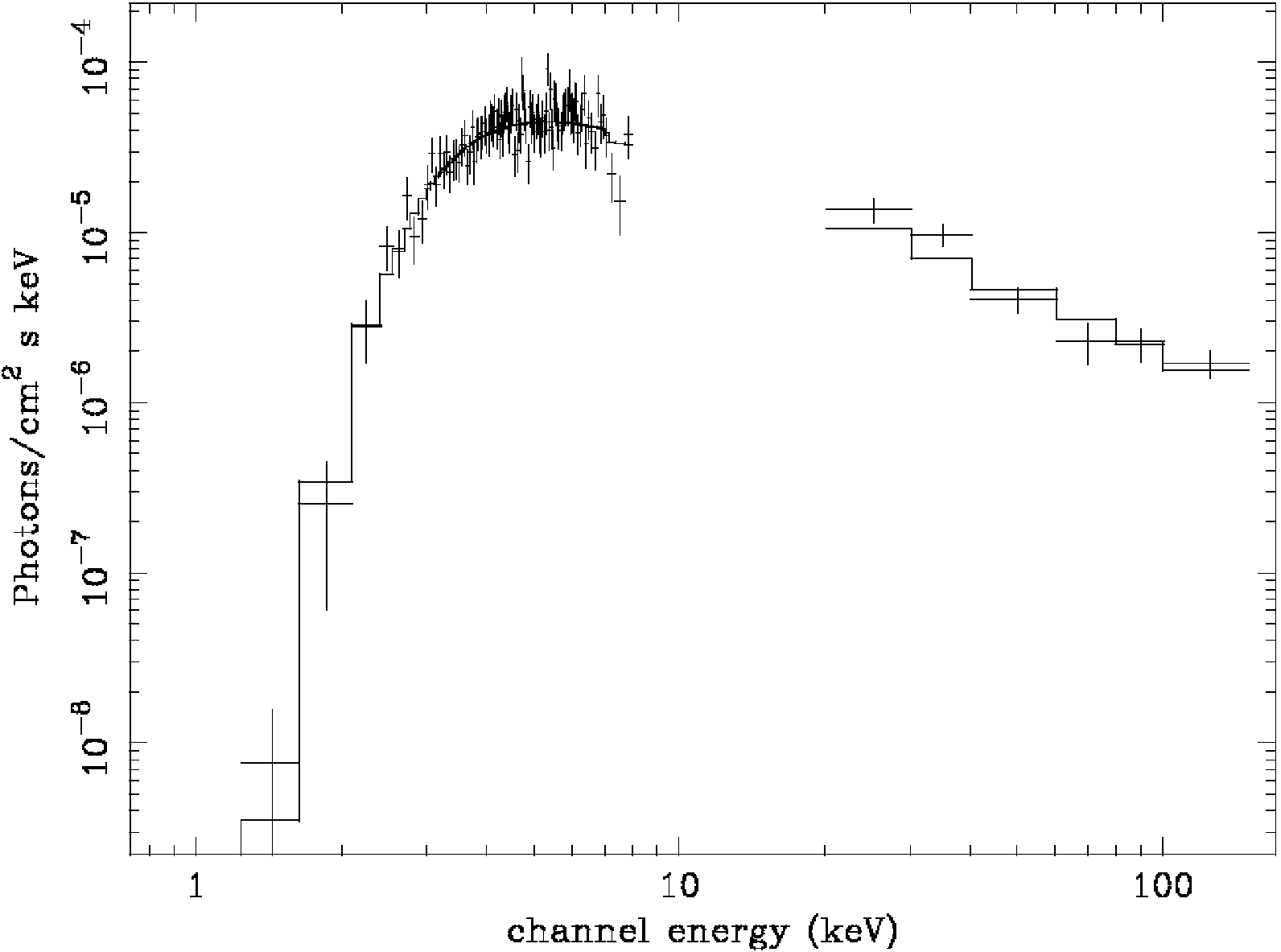} \caption {Combined Chandra-ACIS/IBIS-ISGRI spectrum of
source 3 = NGC6334B. \label{fig4}} 
\end{figure}


\begin{thebibliography}{}

\bibitem{}
Bird, A. J., Barlow, A.J., Bassani, L. et al. 2004, Ap.J. 607, L33
\bibitem{}
Bird, A. J.,Barlow, A.J., Bassani, et al. 2005, Ap.J. in press
\bibitem{}
Cutri, R. M., Skrutskie, M. F., van Dyk, S. et al. 2003 
http://www.ipac.caltech.edu/2mass/releases/second/doc/explsup.html
\bibitem{}
Dickel, H.R., Dickel, J.R., \& Wilson, W.J. 1977,Ap.J. 217, 56
\bibitem{}
Dickey, J.M. \& Lockman, F.J. 1990, A.R.A.A. 28, 215
\bibitem{}
Hofner, P., Delgado, H., Whitney, B., Churchwell, E,. Linz, H. 2002, Ap.J. 579, 95
\bibitem{}
Garmire, G., Feigelson, E. D., Broos, P., Hillenbrand, L. A., Pravdo, S. H., et al. 2000, A.J. 120, 1426 
\bibitem{}
Gehrels, N., Chincarini, G., Giommi, P. et al. 2004, Ap.J. 611, 1005
\bibitem{}
Gros, A., Goldwurm, A., Cadolle-Bel, M., et al. 2003, A\&A 411, L179
\bibitem{}
Hill, J. E., Burrows, D. N.,Nousek, J. A. et al. 2004, SPIE 5165, 217
\bibitem{}
Kohno, M., Koyama, K, and Hamaguchi, K. 2002, Ap.J. 567, 423
\bibitem{}
Kraemer, K. E. \& Jackson, J.M. 1999, Ap.J.S. 124, 439
\bibitem{}
Lebrun, F., Leray, J. P., Lavocat, P., et al. 2003, A\&A 411L, 141
\bibitem{}
McBreen, B., Fazio, G. G., Stier, M., Wright, E. L. 1979,  Ap.J. 232, L183
\bibitem{}
Monet, D. G., Levine, S. E., Canzian, B. et al. 2003,  A.J.  125, 984
\bibitem{}
Moran, J. M., Greene, B., Rodriguez, L. F., Backer, D. C. 1990, Ap.J., 348, 147
\bibitem{}
Neckel, T. 1978 Astr. Ap. 69, 51
\bibitem{}
Revnivtsev, M. G., Churazov, E. M., Sazonov, S. Yu. et al.  2004, A\&A 425, L49
\bibitem{}
Rodriguez, L. F.,  Canto, J. \& Moran, J.M.  1982, Ap.J. 255, 103
\bibitem{}
Sekimoto, Y., Matsuzaki, K., Kamae, T., et al. 2000,P.A.S.J. 51, L31  
\bibitem{}
Straw,S.M., Hayland, A.R, McGregor, P.J. 1989, Ap.J.Suppl.69,99
\bibitem{}
Ubertini P., Lebrun F., Di Cocco G., et al. 2003, A\&A 411, L131 
\bibitem{}

\end{thebibliography}
\end{document}